\documentclass{scienceadvhack}

\usepackage{scicite}
\usepackage{times}
\usepackage{graphicx}

\newcounter{lastnote}

\title{Pinning down the superfluid and measuring masses using pulsar glitches}

\author
{Wynn C.~G. Ho,$^{1\ast}$
Crist\'{o}bal M. Espinoza,$^{2}$
Danai Antonopoulou,$^{1,3}$
Nils Andersson$^{1}$
\\
\\
\normalsize{$^{1}$Mathematical Sciences and STAG Research Centre,
University of Southampton,}
\normalsize{Southampton, SO17 1BJ, UK}\\
\normalsize{$^{2}$Instituto de Astrof\'{i}sica, Facultad de F\'{i}sica,
Pontificia Universidad Cat\'{o}lica de Chile,}
\normalsize{Casilla 306, Santiago 22, Chile}\\
\normalsize{$^{3}$Astronomical Institute Anton Pannekoek, University of
Amsterdam,}
\normalsize{Postbus 94249, NL-1090 GE Amsterdam, Netherlands}\\
\\
\normalsize{$^\ast$Corresponding author. E-mail: wynnho@slac.stanford.edu}
}

\date{}

\begin{document}

\maketitle 

\begin{abstract}
Pulsars are known for their superb timing precision, although glitches can
interrupt the regular timing behavior when the stars are young.
These glitches are thought to be caused by interactions between normal and
superfluid matter in the crust of the star.
However, glitching pulsars such as Vela have been shown to require a
superfluid reservoir that greatly exceeds that available in the crust.  
We examine a model in which glitches tap the superfluid in the core.
We test a variety of theoretical superfluid models against the most recent
glitch data and find that only one model can successfully explain up to
45 years of observational data.
We develop a new technique for combining radio and X-ray data to measure
pulsar masses, thereby demonstrating how current and future telescopes can
probe fundamental physics such as superfluidity near nuclear saturation.
\end{abstract}

\section*{Introduction}
\subsubsection*{Pulsar glitches}
Pulsars are rotating neutron stars born in the collapse and supernova
explosion at the end of a massive star's life.
With a mass larger than that of the Sun and only about 25~km in diameter,
neutron stars are primarily composed of neutron-rich matter near and
above nuclear densities.
Pulsars rotate at incredible speeds, with observed spin periods $P$
ranging from 1.4~ms \cite{hesselsetal06} to more than 1~s.
These rotation periods are very stable, with some rivaling the
precision of atomic clocks.
An array of these distant high-precision clocks is being used in
a global effort to detect gravitational waves from supermassive black
holes at cosmic distances.

The precise timing behavior of many young pulsars is interrupted by
sudden changes, so-called glitches, in their spin period.
Pulsars emit beamed electromagnetic radiation (which is what
allows us to detect a pulsar when its beam crosses our line of sight, like a
lighthouse).  This loss of energy comes at the expense of the pulsar's
rotational energy, causing the pulsar to spin more slowly over time
(characterized by the time derivative of the spin rate $\dot{\Omega}$,
with $\Omega=2\pi/P$).
However, during glitches, the pulsar spin rate suddenly
increases over a very short time ($<30\mbox{ s}$) \cite{dodsonetal07}
and usually relaxes to
its pre-glitch rate over a longer time (tens to hundreds of days).
Examples of the accumulated effect of glitches on the spin rate of two pulsars,
Vela (or PSR~B0833$-$45) and PSR~J0537$-$6910, are given in Fig.~1,
where each step-like increase of spin rate is a glitch.

\begin{figure}[!b]
\begin{center}
\includegraphics[scale=0.21]{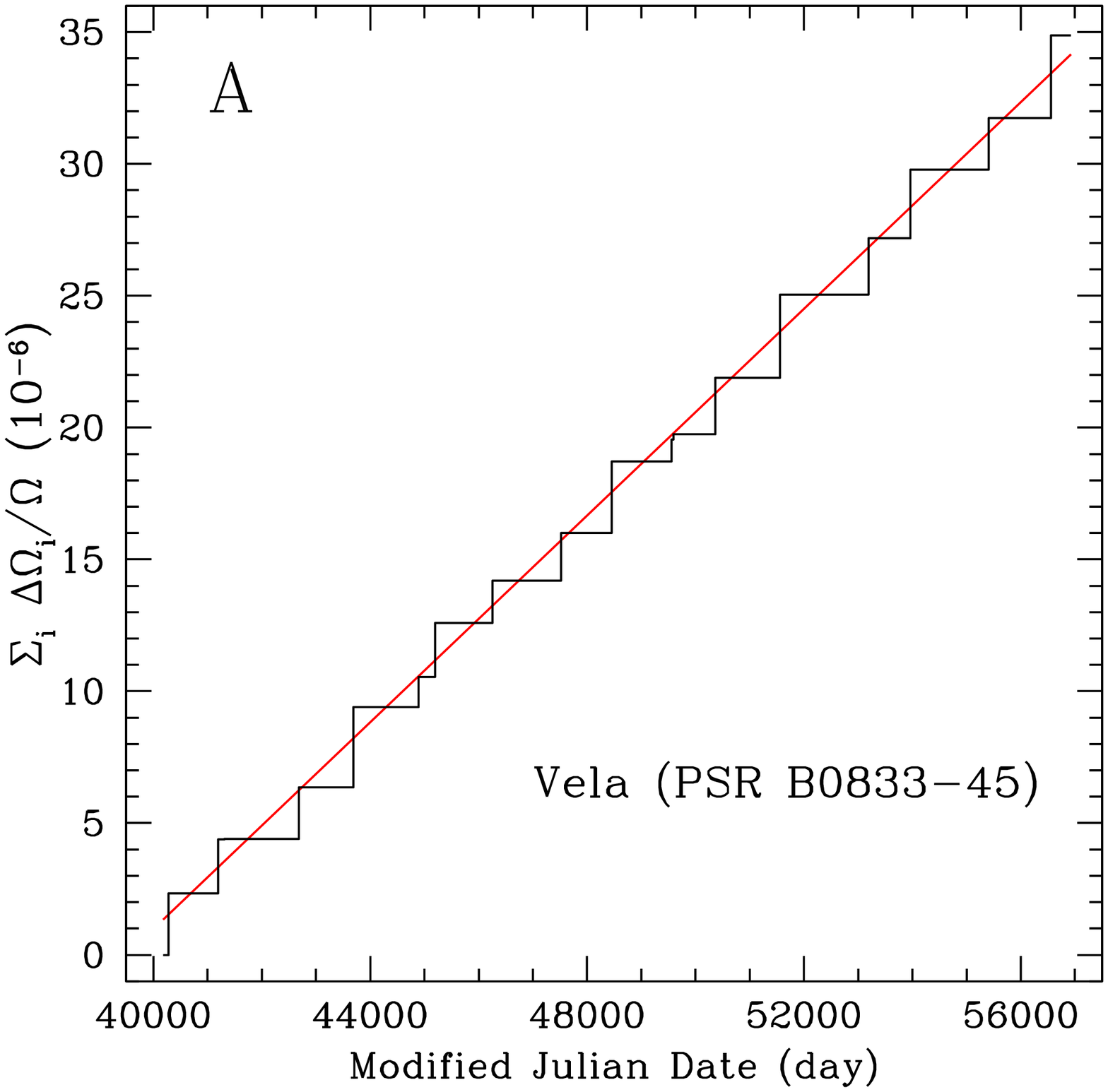}
\includegraphics[scale=0.21]{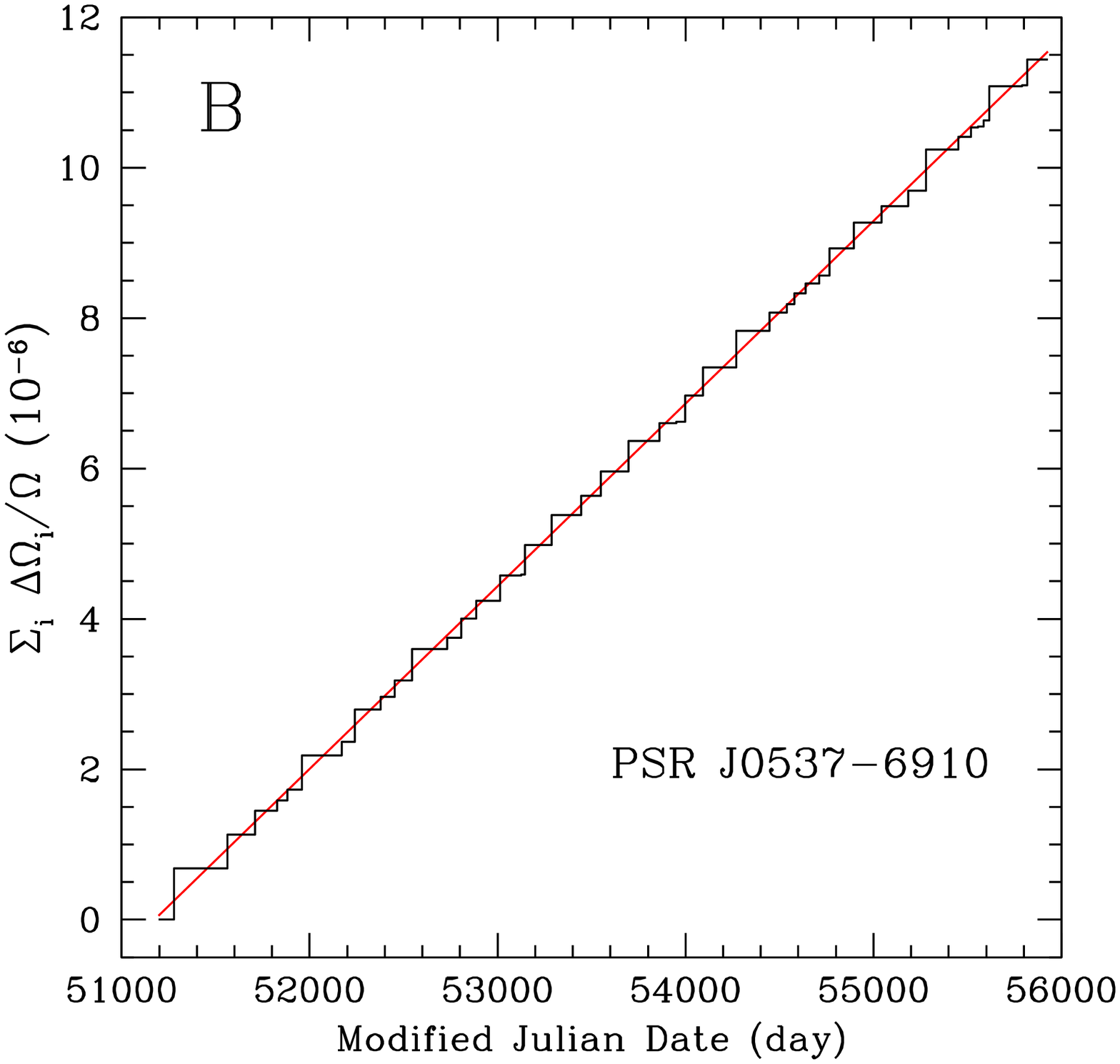}
\caption{
\noindent {\bf Cumulative fractional change of pulsar spin
frequency at glitches over time. (A) Vela. (B) PSR~J0537$-$6910.}
Glitch data for Vela are from \cite{jodrell},
whereas data for PSR~J0537$-$6910 are from \cite{middleditchetal06}
plus 22 new events found in \cite{antonopoulouetal15}.
Straight lines are a least-squares fit, with the average activity parameter
$\langle A\rangle$ as the slope.
}
\end{center}
\end{figure}

Glitches are believed to be the manifestation of a neutron superfluid in the
inner crust of a pulsar.
The structure of a neutron star can be divided into three regions:
the outer crust, inner crust, and core.
The outer crust is composed of a crystalline solid of normal matter at
densities below neutron drip
(at mass density $\sim 4\times 10^{11}\mbox{ g cm$^{-3}$}$).
Above approximately half the nuclear saturation baryon density
of $n_{\rm b}\sim 0.16\mbox{ fm$^{-3}$}$,
the core is predominantly composed of neutrons in a liquid state.
Between the outer crust and core, the inner crust contains
neutron-rich nuclei embedded in a sea of free neutrons.
These free neutrons are expected to be in a superfluid state because the
critical temperature $T_{\rm c}$ (below which neutrons become superfluid)
is well above the typical temperature of pulsars ($<10^9\mbox{ K}$).

\subsubsection*{Superfluid in the neutron star crust}
Unlike normal matter (for example, that in the outer crust), superfluid matter
in the inner crust rotates by forming vortices whose areal density determines
the spin rate of the superfluid.
To decrease its spin rate, superfluid vortices must move
so that the areal density decreases.
In the inner crust of a neutron star, these vortices are usually pinned
to the nuclei of normal matter \cite{andersonitoh75}.
While the rest of the star slows down owing to electromagnetic energy loss,
the neutron superfluid does not.
As a result, this superfluid can act as a reservoir of angular momentum.
Over many pulsar rotations,
an increasing lag develops between the stellar spin rate and that
of the neutron superfluid in the inner crust.
When this lag exceeds a critical (but unknown) value, superfluid vortices 
unpin and transfer their angular momentum to the rest of the star, causing
the stellar rotation rate to increase and producing what we observe as a
glitch \cite{baymetal69,andersonitoh75}.
Link {\it et~al.} \cite{linketal99} calculated both the minimum angular momentum
needed to produce the glitches that are observed in seven pulsars and
the crust moment of inertia predicted by theoretical models of neutron stars.
They find that the theoretical moment of inertia matches that required by
glitch observations,
thus strongly supporting the above glitch model and providing key evidence for
the superfluid component in the stellar crust.
In particular, the superfluid reservoir must exceed
the observable quantity $G\equiv 2\tau_{\rm c}\langle A\rangle$,
where $\tau_{\rm c}=\Omega/2\dot{\Omega}$ is the pulsar characteristic age,
$\langle A\rangle=(1/t_{\rm obs})\sum\Delta\Omega/\Omega$ is the
average activity parameter \cite{mckennalyne90},
and $t_{\rm obs}$ is the time span over which the pulsar has been observed
(years to decades in the cases studied here).
The average activity $\langle A\rangle$ can be determined by the slope of a
linear fit to glitches, as shown by the solid lines in Fig.~1,
and results in $G=1.62\pm0.03\%$ for Vela and
$G=0.875\pm0.005\%$ for PSR~J0537$-$6910.
These values of $G$ should be compared to predictions of the crust moment
of inertia relative to the total moment of inertia $I$ of $\approx 3-5\%$
at most [for the characteristic neutron star mass of $1.4\,M_{\rm Sun}$;
see, for example, \cite{anderssonetal12,piekarewiczetal14}],
depending on the theoretical model of the nuclear equation of state (EOS).
Furthermore, the regularity of large, similarly sized glitches from Vela and
PSR~J0537$-$6910 implies that glitches in these pulsars are tapping and
essentially exhausting the entire superfluid reservoir rather than a small
fraction of it.

\subsubsection*{Recent advances}
Since the work of Link {\it et~al.} \cite{linketal99}, much progress has
been made in understanding superfluidity in neutron star crusts.
In particular, Chamel \cite{chamel05,chamel12} found that the effect of
entrainment makes it very difficult to move superfluid neutrons relative
to the crust lattice.
As a result, Andersson {\it et~al.} \cite{anderssonetal12} and Chamel
\cite{chamel13} found that the previous calculations of Link {\it et~al.}
\cite{linketal99} underestimate, by a factor of $\approx 4.3$,
the moment of inertia required by observed glitches and that the
superfluid reservoir in the crust of neutron stars is
insufficient to produce the observed size and frequency of glitches.
For example, using the most up-to-date data
\cite{jodrell,kuiperhermsen15,antonopoulouetal15},
19 glitches seen during 45 years of observing the Vela
pulsar require a reservoir comprising $4.3\times1.6\%=6.9\%$
of the total moment of inertia,
whereas 45 glitches seen during 13 years for PSR~J0537$-$6910 require
$4.3\times0.9\%=3.9\%$ of the total.

More recently, Piekarewicz {\it et~al.} \cite{piekarewiczetal14} and
Steiner {\it et~al.} \cite{steineretal15} have shown that there is
sufficient uncertainty in the theoretical nuclear EOS
that determines the size of neutron star crusts for the crust to
have enough moment of inertia to explain the glitches.
Although possible, this argument does not take into account superfluidity, and
more importantly, these authors \cite{piekarewiczetal14,steineretal15}
note that their solution is in conflict with other observations.
The proposed EOSs that meet the glitch requirement predict neutron star
radii of $\approx 14\pm 0.5\mbox{ km}$ for typical neutron star masses of
$1.2$ to $2M_{\rm Sun}$, which is in contrast to the observationally inferred
radii of $\approx 11.8\pm 0.9\mbox{ km}$ \cite{lattimersteiner14}
or even smaller \cite{guillotetal13}.
Here, we include a large number of more recent glitch data
from \cite{jodrell,yuetal13,antonopoulouetal15} and
explore a solution proposed in \cite{anderssonetal12}.
In doing so, we find an unexpected and rather remarkable result.

\section*{Results}
\subsubsection*{Temperature dependence of superfluid}
In the previous analyses
\cite{linketal99,anderssonetal12,chamel13,piekarewiczetal14,steineretal15},
pulsar glitches are assumed to tap the angular momentum reservoir
associated with superfluid neutrons in the inner crust of the star
(although the temperature dependence of superfluidity is largely ignored).
Therefore, by calculating the entire moment of inertia of the crust,
it is possible to determine the maximum reservoir available for producing
glitches, and this is found to be smaller than that needed to explain
observed glitch activity \cite{anderssonetal12,chamel13}, unless the
crust is unusually thick \cite{piekarewiczetal14,steineretal15}.
Here, we consider a superfluid reservoir that extends into the stellar core
and account for the temperature dependence of superfluidity.
Importantly, we use the latter in comparison with the observed temperature
of pulsars.
Figure~2 shows several example models of the critical temperature of
neutron (singlet-state) superfluidity as a function of
$n_{\rm b}$ for neutron stars built using the BSk20 EOS \cite{potekhinetal13}. 
The boundary between the crust (shaded region) and the core is at baryon density
$n_{\rm b}=0.0854\mbox{ fm$^{-3}$}$
and is denoted by the vertical solid line.
Because the correct nuclear EOS is unknown, we also consider the
BSk21 EOS \cite{potekhinetal13} and APR EOS \cite{akmaletal98}.
All three EOS models produce neutron stars with radius
$R\approx 11$ to $12.5\mbox{ km}$, thus satisfying the observational constraint
from \cite{lattimersteiner14}, and maximum mass greater than the highest
observed masses \cite{demorestetal10,antoniadisetal13}.
The superfluid models we use are parameterized fits to nuclear physics
calculations; see \cite{hoetal15} and references therein for details.
For most superfluid models, the critical temperature, and hence allowed region
for neutrons to become superfluid, is confined to the inner crust,
that is, in the shaded region to the left of the vertical solid line.
Therefore, pulsar glitches can only involve the moment of inertia
of the inner crust if one of these superfluid models is the correct one.
However, there are a few superfluid models that extend into the core,
for example,
the solid curve labeled SFB, which is the model from \cite{schwenketal03}.
For superfluid models such as the SFB model, if the temperature of a pulsar is
low enough so that neutrons in the inner crust {\it and} outer core are
superfluid, then glitches from the pulsar could involve additional moment
of inertia from the core.

\begin{figure}
\begin{center}
\includegraphics[scale=0.42]{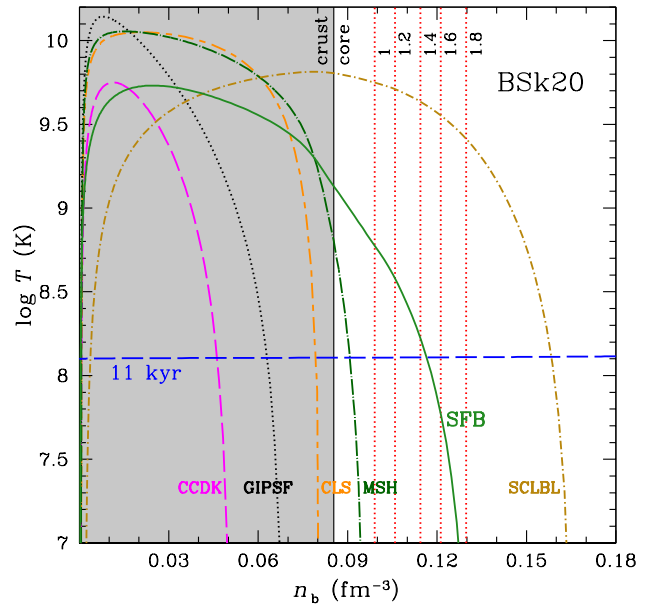}
\caption{
\noindent {\bf Temperature dependence of neutron superfluid models
as a function of baryon number density for the BSk20 nuclear equation of
state.}
Thick curved lines are the superfluid critical temperature for the (labeled)
models from \cite{hoetal15}.
The vertical solid line indicates the separation between the crust
(shaded region) and the core.
Vertical dotted lines denote the density at which the superfluid moment of
inertia (using the SFB superfluid model) is 1.6\% of the total stellar moment
of inertia for neutron stars of different mass (labeled in units of solar mass).
The (nearly horizontal) dashed line is the temperature of a $1.4\,M_{\rm Sun}$
neutron star at an age of 11000 years.
}
\end{center}
\end{figure}

We compute the total and partial moments of inertia
following the method described in \cite{anderssonetal12}.
Figure~2 shows our findings for the BSk20 EOS;
BSk21 and APR EOSs produce qualitatively similar results.
The vertical dotted lines in Fig.~2 indicate the density at which the
moment of inertia of the superfluid (using the SFB superfluid model) exceeds
$G = 1.6\%$ (from the Vela pulsar) for neutron stars of various mass.
Low mass neutron stars have a thicker crust than high mass stars, and
we see that a neutron superfluid confined to the crust can only provide
a relative moment of inertia of 1.6\% for neutron stars with a mass much
less than $1\,M_{\rm Sun}$ for BSk20.
All glitching pulsars with $G\approx 1\%$ would have to be of low mass
($<1\,M_{\rm Sun}$) as well.
For typical neutron star masses of $1.2$ to $2\,M_{\rm Sun}$ \cite{lattimer14},
some small fraction of the core must contribute to the moment of inertia
required by glitches seen in, for example, the Vela pulsar.

The next natural question is whether enough neutrons in this moment of
inertia are actually superfluid,
that is in Fig.~2, to the left of one of the vertical dotted lines and below
the superfluid critical temperature $T_{\rm c}$.
To answer this question, we need to determine the interior temperature
$T(n_{\rm b})$ of a neutron star and evaluate at what
densities $n_{\rm b}$ the inequality $T<T_{\rm c}$ is satisfied.
This will vary for each
pulsar, depending on its age and/or measured temperature;
for example, the Vela pulsar is $11000^{+5000}_{-5600}\mbox{ years}$ old
\cite{pageetal09,tsurutaetal09}
and has a surface temperature of $9.3\times 10^5\mbox{ K}$
\cite{viganoetal13},
whereas PSR~J0537$-$6910 only has an age determination of
$2000^{+3000}_{-1000}\mbox{ years}$ \cite{wanggotthelf98,chenetal06}.
Neutron stars are born in supernovae at very high temperatures, but they
cool rapidly because of efficient emission of neutrinos.
We perform neutron star cooling simulations using standard (``minimal'')
neutrino emission processes \cite{yakovlevpethick04,pageetal06}
to find the interior temperature at various ages;
see \cite{hoetal12} for details.
Figure~2 plots the resulting temperature profile
(at the age of the Vela pulsar; dashed line)
for a $1.4\,M_{\rm Sun}$ neutron star built using the BSk20 EOS.
The temperature profile is different for different mass but not
drastically so, unless neutrino emission by (non-minimal) direct Urca
processes occurs (see below).
We find that,
among nine superfluid models that span a wide range in parameter space
[see \cite{hoetal15} for references],
only the SFB model provides a superfluid reservoir of the required level
(with a maximum relative moment of inertia of 2.9\% for an old and cold
$1.4\,M_{\rm Sun}$ neutron star).
For superfluid models that are confined to the crust, the reservoir is too
small, whereas the reservoir is too large for models that extend much deeper
into the core.  The latter would be unable to explain the regularity of
similar-sized glitches, which requires the reservoir to be completely
exhausted in each event.
The near intersection of the three lines
[vertical dotted line for glitch requiring $G=1.6\%$ at $1.4\,M_{\rm Sun}$,
solid line for the SFB model of superfluid critical temperature $T_{\rm c}$,
and horizontal dashed line for neutron star temperature $T(n_{\rm b})$ at age
$=11000\mbox{ years}$] is one of our key findings:
{\it The mass of the Vela pulsar is near the characteristic value of
$1.4\,M_{\rm Sun}$,
and the size and frequency of Vela's observed glitches are a natural
consequence of the superfluid moment of inertia available to it at its
current age}.

\subsubsection*{Pulsar mass from glitches}
We summarize our results in Fig.~3,
which shows the interior temperature $T$ of a pulsar as a function of the
value of $G=2\tau_{\rm c}\langle A\rangle$.
These two quantities, $G$ and $T$, are directly determined from observational
data.
The former comes from (radio or X-ray) observations of glitches
and is best determined for pulsars that undergo regular large glitches.
The latter is obtained either from the age of the pulsar [for example,
by determining the age of an associated supernova remnant (SNR)] or by
measuring/constraining the
surface temperature of the pulsar through X-ray measurements
(see Materials and Methods).
The age gives the interior temperature via neutron star cooling simulations
such as the ones conducted here, whereas the surface temperature is related to
the interior temperature via well-known relationships \cite{potekhinetal03},
which depend on the composition of the outer layers of the star (taken
to be iron here).
Thus for a given pulsar that has these two measurable quantities,
this figure allows us to determine the pulsar's mass.
For example, using the BSk20 EOS, we find that
Vela is a $1.51\pm0.04\,M_{\rm Sun}$ neutron star and
PSR~J0537$-$6910 is a $1.83\pm0.04\,M_{\rm Sun}$ neutron star.
These two represent our two best cases because they have the smallest
uncertainty in $G$ and relatively small uncertainty in $T$
(see Fig.~3 and Table~1).

\begin{figure}
\begin{center}
\includegraphics[scale=0.42]{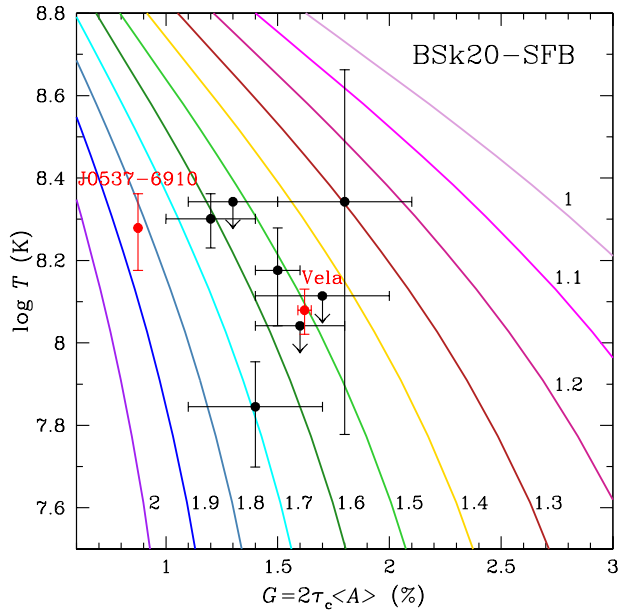}
\caption{
\noindent {\bf Neutron star mass from pulsar observables $G$ and
interior temperature $T$.}
Data points are for pulsars with measured $G$ from glitches and $T$
from an age or surface temperature observation (see Table~1).
Lines (labeled by neutron star mass, in units of solar mass) are the
theoretical prediction for $G$ and $T$ using the BSk20 nuclear equation of
state and SFB neutron superfluid models.
}
\end{center}
\end{figure}

Figure~3 plots the data for other glitching pulsars (see also Tables 1 and 2).
The nine sources shown are currently the
best pulsars for our prescribed technique of measuring neutron star mass.
They are selected based on each undergoing several glitches that
are approximately regularly spaced in time and similarly sized,
so that a linear fit such as the ones illustrated in Fig.~1 is a good
characterization of their glitch behavior.
Note that there are likely at least two types of glitches
\cite{espinozaetal11,yuetal13}, and
glitching pulsars with $G\ll 1$ would only be tapping a small portion of the
superfluid moment of inertia (incidentally, this is the reason for excluding
the Crab pulsar).  Good candidates for this technique must also have an age
or temperature constraint.

\begin{table*}[htb]
\begin{center}
\caption{\bf Characteristics of glitching pulsars}
\small{
\begin{tabular}{|lcccclclc|}
\hline
PSR & $\tau_{\rm c}$ & $\langle A\rangle$ & $G$ & Age & Ref. & $T_{\rm s}^\infty$ & Ref. & $T$ \\
& ($10^3$~yr) & ($10^{-9}\mbox{ d$^{-1}$}$) & (\%) & ($10^3$~yr) & & ($10^6$~K) & & ($10^8$~K) \\ \hline
J0537$-$6910 & 4.93 & $2.43\pm0.01$ & $0.875\pm0.005$ & 1--5 & \cite{wanggotthelf98,chenetal06} & & & $1.9\pm0.4$ \\
B0833$-$45 & 11.3 & $1.96\pm0.04$ & $1.62\pm0.03$ & 5.4--16 & \cite{pageetal09,tsurutaetal09} & $0.93^{\rm a}$ & \cite{viganoetal13} & $1.2\pm0.15$ \\
B1046$-$58 & 20.4 & $1.1\pm0.2$ & $1.6\pm0.2$ & & & $<1.4^{\rm b}$ & \cite{gonzalezetal06} & $<1.1$ \\
B1338$-$62 & 12.1 & $1.7\pm0.1$ & $1.5\pm0.1$ & 3--30 & \cite{caswelletal92} & & & $1.5\pm0.4$ \\
B1706$-$44 & 17.8 & $1.1\pm0.2$ & $1.4\pm0.3$ & 5--18 & \cite{abramowskietal11} & 0.5--$0.8^{\rm a}$ & \cite{mcgowanetal04} & $0.7\pm0.2$ \\
B1757$-$24 & 15.5 & $1.5\pm0.2$ & $1.7\pm0.3$ & $>10$ & \cite{blazeketal06,zeigeretal08} & & \cite{kaspietal01} & $<1.3$ \\
B1800$-$21 & 15.8 & $1.5\pm0.2$ & $1.8\pm0.3$ & & & 1--$3^{\rm b}$ & \cite{kargaltsevetal07} & $2.2^{+2.4}_{-1.6}$ \\
B1823$-$13 & 21.5 & $0.8\pm0.1$ & $1.3\pm0.2$ & & & $<2^{\rm b}$ & \cite{pavlovetal08} & $<2.2$ \\
1E~1841$-$045 & 4.57 & $4\pm1$ & $1.2\pm0.2$ & 0.75--2.1 & \cite{kumaretal14} & $3^{\rm b}$ & \cite{viganoetal13} & $2\pm0.3$ \\
\hline
\multicolumn{9}{l}{$^{\rm a}$ From an atmosphere spectral model.
\quad $^{\rm b}$ From a blackbody spectral model.}
\end{tabular}
}
\end{center}
\end{table*}

\begin{table}[!h]
\begin{center}
\caption{\bf Mass, radius, and total moment of inertia of glitching pulsars for
three nuclear equation of state models}
\begin{tabular}{|lccc|}
\hline
PSR & $M$ & $R$ & $I$ \\
& ($M_{\rm Sun}$) & (km) & ($M_{\rm Sun}$ km$^2$) \\ \hline
 \multicolumn{4}{|c|}{BSk20 EOS} \\
J0537$-$6910 & $1.83_{-0.04}^{+0.04}$ & $11.5_{-0.05}^{+0.05}$ & $81.4_{-1.6}^{+1.5}$ \\
B0833$-$45 & $1.51_{-0.04}^{+0.04}$ & $11.7_{-0.02}^{+0.02}$ & $66.7_{-2.0}^{+2.0}$ \\
B1046$-$58 & $1.53_{-0.08}^{+0.36}$ & $11.7_{-0.3}^{+0.03}$ & $67.7_{-4.0}^{+16.0}$ \\
B1338$-$62 & $1.52_{-0.10}^{+0.11}$ & $11.7_{-0.06}^{+0.03}$ & $67.2_{-5.1}^{+5.4}$ \\
B1706$-$44 & $1.69_{-0.17}^{+0.19}$ & $11.6_{-0.2}^{+0.1}$ & $75.4_{-8.2}^{+7.9}$ \\
B1757$-$24 & $1.46_{-0.12}^{+0.43}$ & $11.7_{-0.3}^{+0.03}$ & $64.1_{-6.2}^{+19.5}$ \\
B1800$-$21 & $1.30_{-0.35}^{+0.37}$ & $11.8_{-0.1}^{+0.01}$ & $55.8_{-18.3}^{+18.7}$ \\
B1823$-$13 & $1.53_{-0.10}^{+0.58}$ & $11.7_{-0.9}^{+0.03}$ & $67.7_{-5.1}^{+19.8}$ \\
1E~1841$-$045 & $1.61_{-0.14}^{+0.16}$ & $11.7_{-0.1}^{+0.06}$ & $71.6_{-7.0}^{+7.3}$ \\
\hline \multicolumn{4}{|c|}{BSk21 EOS} \\
J0537$-$6910 & $2.11_{-0.05}^{+0.04}$ & $12.1_{-0.1}^{+0.1}$ & $106.8_{-1.3}^{+0.7}$ \\
B0833$-$45 & $1.82_{-0.04}^{+0.04}$ & $12.5_{-0.03}^{+0.03}$ & $94.9_{-2.1}^{+2.1}$ \\
B1046$-$58 & $1.85_{-0.09}^{+0.35}$ & $12.5_{-0.7}^{+0.06}$ & $96.5_{-4.8}^{+11.0}$ \\
B1338$-$62 & $1.82_{-0.11}^{+0.12}$ & $12.5_{-0.1}^{+0.06}$ & $94.9_{-6.1}^{+5.9}$ \\
B1706$-$44 & $2.01_{-0.16}^{+0.16}$ & $12.3_{-0.4}^{+0.2}$ & $103.8_{-7.3}^{+3.9}$ \\
B1757$-$24 & $1.77_{-0.13}^{+0.43}$ & $12.5_{-0.7}^{+0.05}$ & $92.2_{-7.4}^{+15.2}$ \\
B1800$-$21 & $1.56_{-0.42}^{+0.44}$ & $12.6_{-0.3}^{+0.01}$ & $80.0_{-27.5}^{+23.3}$ \\
B1823$-$13 & $\ge1.71$ & $\le12.6$ & $\ge88.9$ \\
1E~1841$-$045 & $1.90_{-0.15}^{+0.16}$ & $12.4_{-0.2}^{+0.1}$ & $99.0_{-7.8}^{+6.5}$ \\
\hline \multicolumn{4}{|c|}{APR EOS} \\
J0537$-$6910 & $2.05_{-0.03}^{+0.04}$ & $10.9_{-0.1}^{+0.07}$ & $86.4_{-0.7}^{+0.7}$ \\
B0833$-$45 & $1.80_{-0.03}^{+0.03}$ & $11.3_{-0.03}^{+0.03}$ & $77.6_{-1.3}^{+1.3}$ \\
B1046$-$58 & $1.82_{-0.07}^{+0.29}$ & $11.3_{-0.5}^{+0.07}$ & $78.4_{-3.1}^{+8.9}$ \\
B1338$-$62 & $1.81_{-0.09}^{+0.09}$ & $11.3_{-0.1}^{+0.09}$ & $78.0_{-4.0}^{+3.7}$ \\
B1706$-$44 & $1.96_{-0.14}^{+0.14}$ & $11.1_{-0.3}^{+0.2}$ & $83.8_{-5.4}^{+3.4}$ \\
B1757$-$24 & $1.76_{-0.11}^{+0.35}$ & $11.4_{-0.6}^{+0.09}$ & $75.8_{-5.1}^{+11.5}$ \\
B1800$-$21 & $1.60_{-0.33}^{+0.33}$ & $11.5_{-0.3}^{+0.2}$ & $68.4_{-15.9}^{+14.4}$ \\
B1823$-$13 & $1.82_{-0.09}^{+0.36}$ & $11.3_{-0.9}^{+0.09}$ & $78.4_{-4.0}^{+7.8}$ \\
1E~1841$-$045 & $1.89_{-0.13}^{+0.12}$ & $11.2_{-0.2}^{+0.1}$ & $81.3_{-5.5}^{+4.1}$ \\
\hline
\end{tabular}
\end{center}
\end{table}

\section*{Conclusion\\ \\}
We conclude by commenting on the nuclear EOS and superfluid models
that are adopted and that lead to the results presented here.
For the former, we focus on and describe the results for BSk20.
The other two EOSs we consider, BSk21 and APR, yield qualitatively similar
results as BSk20, except the mass of most pulsars is about $1.8\,M_{\rm Sun}$,
compared to $1.5\,M_{\rm Sun}$ for BSk20, and the mass of PSR~J0537$-$6910 is
$> 2\,M_{\rm Sun}$ (see Table~2).
Such high mass pulsars (for BSk21 and APR) would cool to a much lower
temperature than observed because fast (direct Urca) neutrino emission processes
become operative above $1.59\,M_{\rm Sun}$ for BSk21 and above
$1.96\,M_{\rm Sun}$ for APR.
Note that direct Urca processes do not take place for any mass using BSk20.
Uncertainty in the EOS is the primary contributor to the systematic error in
our mass determinations.
An indication of the effect of this uncertainty can be seen in Table~2:
errors for $M$, $R$, and $I$ arise from the errors in $G$ and $T$
(see Fig.~3), in contrast to the difference in a particular parameter for
different EOS models.
For neutron superfluidity,
we concern ourselves only with a superfluid in the
singlet-state ($^1$S$_0$), which is likely present in the inner crust and
possibly in a small fraction of the outer core of a neutron star.
Deeper within the star, neutrons can be in a superfluid triplet-state
($^3$P$_2$).
The properties of this superfluid are much more uncertain.
Recent measurements of the rapid cooling of a young neutron star in the
Cassiopeia~A supernova remnant have revealed the critical temperature of
the triplet-state neutron superfluid \cite{pageetal11,shterninetal11},
although this finding is currently under debate
\cite{elshamoutyetal13,posseltetal13}.
The peak critical temperature of the triplet-state is much lower than that
of the singlet-state, although there may be an overlap region, and the
superfluid properties are unclear within this region.

The possibility of measuring the mass of isolated pulsars has not been
previously demonstrated.  To date, the most precise neutron star mass
measurements are by radio timing of pulsars that are in a binary star
system \cite{lattimer14},
for example, neutron stars with the highest measured mass of
$1.97\pm0.04\,M_{\rm Sun}$ \cite{demorestetal10}
and $2.01\pm0.04\,M_{\rm Sun}$ \cite{antoniadisetal13}.
The number of pulsars that are seen to glitch continues to increase
\cite{espinozaetal11,yuetal13}, along with ongoing discoveries of pulsars,
including the binary system with PSR~J2032+4127 \cite{lyneetal15}, which
could be used to test our method in the future.
Our method of measuring neutron star masses can greatly increase the
number of known masses, thereby allowing the determination of fundamental
physics properties such as the nuclear EOS and superfluidity.
Although there are currently relatively large systematic uncertainties,
these will improve as our knowledge of the physics of dense matter improves.
The novelty of our approach is the combination of pulsar glitch data and
the temperature dependence of superfluidity.
The method is especially promising with upcoming large astronomical
observatories such as the Square Kilometer Array (SKA) in radio and Athena+
in X-rays.
SKA could discover all the observable pulsars in the Galaxy, and we show
that a program to monitor glitching pulsars could greatly transform the
fields of neutron star and nuclear physics.

\section*{Materials and Methods\\ \\}
Figure~3 and Table~1 present the data for the glitching pulsars studied in
this work.  Here, we explain how this information was determined.
The average activity parameter $\langle A\rangle$ is obtained by a
least-squares fit to the cumulative fractional change of pulsar spin
frequency over time, examples of which are shown in Fig.~1.
The errors for $\langle A\rangle$ given in Table~1 come
directly from the fit to each set of pulsar glitches.  We performed fits to
four subsets of Vela glitches and compared the results to those obtained
using the entire data set.  We find that errors derived from using the
subsets well represent the uncertainty arising from using fewer glitches
due to lack of long-term information.
Age is that of a SNR associated with the
pulsar, except in the cases of PSR~B1706$-$44 (where the association with
SNR G343.1$-$2.3 is uncertain, so we use the SNR age for the lower limit
and the characteristic age for the upper limit)
and PSR~B1757$-$24 (where the association with SNR G5.4$-$1.2 is uncertain,
so we estimate a lower limit that is comparable to its characteristic age).
Surface temperature $T_{\rm s}^\infty$ is obtained from X-ray spectral
constraint/measurement and is the redshifted value, that is, measured by
a distant observer.
Superscript ``a'' indicates that $T_{\rm s}^\infty$ is from fitting the
spectrum with a neutron star atmosphere model,
whereas ``b'' indicates that $T_{\rm s}^\infty$ is from fitting
(or obtaining an upper limit) with a blackbody model;
the latter generally overestimates surface temperature by a factor of
$\sim 1.5$ \cite{potekhin14}.

Neutron stars are nearly isothermal after several hundred years
\cite{yakovlevpethick04,pageetal06}.
Therefore, the temperature at the bottom of the neutron star envelope
(at $\sim 10^{10}\mbox{ g cm$^{-3}$}$) is essentially the temperature in
the deeper crust and core.
To determine the interior temperature $T$ from the age or surface temperature,
we use the following procedure.
If the pulsar only has an age determination, we perform neutron star cooling
simulations \cite{hoetal12} and extract the interior temperature at the
appropriate age, with errors estimated from varying model parameters such
as the mass.
If the pulsar only has a surface temperature measurement, then we convert
surface temperature into interior temperature using the relation given in,
for example, \cite{potekhinetal03}, assuming an iron
envelope composition \cite{changbildsten04}.
If the pulsar has both age and surface temperature measurements, then we use
the age method described above and we verify that the inferred interior
temperature produces a surface temperature that approximately matches the
observed value.
In the case of 1E~1841$-$045, this pulsar is a magnetar, and magnetars are
unusually hot for their age [likely due to magnetic field decay and heating
in the outer layers of the crust; \cite{hoetal12}];
because we are interested in the temperature near the crust-core boundary,
we only use the age to determine $T$ for 1E~1841$-$045.

\bibliography{scibib}

\bibliographystyle{Science}


\paragraph*{Acknowledgments:}
We are grateful to A. Lyne and B. Stappers for unpublished
radio data and L. Kuiper for X-ray ToAs of PSR~J0537$-$6910.
{\bf Funding:}
W.C.G.H., D.A., and N.A. acknowledge support from STFC in the UK.
C.M.E acknowledges support from FONDECYT (postdoctorado 3130152).
D.A. acknowledges support from a NWO Aspasia grant
(principal investigator: A.L. Watts) and
the research networking programme NewCompStar (COST Action MP1304).
{\bf Author contributions:}
W.C.G.H. contributed to developing the model, performed the model calculations,
and wrote the manuscript.
C.M.E. gathered and analyzed the data and contributed to writing the manuscript.
D.A. gathered and analyzed the data and contributed to writing the manuscript.
N.A. contributed to developing the model and writing the manuscript.
{\bf Competing Interests:}
The authors declare that they have no competing interests.
{\bf Data and materials availability:}
All data presented in this work are available upon request to C.M.E.

\vspace{0.3cm}
\noindent
Submitted 8 May 2015 \\
Accepted 30 July 2015 \\
Published 2 October 2015 \\
10.1126/sciadv.1500578

\paragraph*{Citation:}
W.~C.~G.~Ho, C.~M.~Espinoza, D.~Antonopoulou, N.~Andersson, Pinning down the
superfluid and measuring masses using pulsar glitches. {\it Sci. Adv.}
{\bf 1}, e1500578 (2015).

\end{document}